\documentclass{elsart}
\usepackage{graphicx}
\usepackage{amsmath}
\usepackage{amssymb}
\newcommand{\bq}{\begin{equation}}
\newcommand{\ee}{\end{equation}}
\newcommand{\fr}[2]{\frac{#1}{#2}}

\begin{document}

\begin{frontmatter}



\title{Chaos beyond linearized stability analysis: folding of the phase
space and distribution of Lyapunov exponents}


\author[add1,add2]{P.~G. Silvestrov}
\and
\author[add3,add4]{I.~V. Ponomarev}
\address[add1]{Instituut-Lorentz, Universiteit Leiden, P.O. Box
9506, 2300 RA Leiden, The Netherlands}
\address[add2]{Theoretische Physik III, Ruhr-Universitt Bochum, 44780
Bochum, Germany}
\address[add3]{Naval Research Lab, Washington, DC 20375, USA}
\address[add4]{Queens College of the City University of New
York, Flushing, NY 11367, USA}
%\address{}

\begin{abstract}
We consider a mechanism for area preserving Hamiltonian systems
which leads to the enhanced probability, $P(\lambda ,t)$, to find
small values of the finite time Lyapunov exponent, $\lambda$. In our
investigation of chaotic dynamical systems we go beyond the
linearized stability analysis of nearby divergent trajectories and
consider folding of the phase space in the course of chaotic
evolution. We show that the spectrum of the Lyapunov exponents
$F(\lambda)= \lim_{t\rightarrow\infty} t^{-1}\ln P(\lambda ,t)$ at
the origin has a finite value $F(0)=-\tilde{\lambda}$ and a slope
$F'(0)\le 1$. This means that all negative moments of the
distribution $\langle e^{-m\lambda t}\rangle$ are saturated by rare
events with $\lambda\rightarrow 0$. Extensive numerical simulations
confirm our findings.

\end{abstract}

\end{frontmatter}
% main text
%\section{}
%\label{}
% \appendix
% \section{}
% \label{}

An exponential divergency of nearby trajectories in phase space is
commonly considered as a paradigm of classical
chaos~\cite{Chirikov79,ott}. The time evolution of the distance
between two trajectories is determined by the stability matrix
${\cal M}(t)$, whose largest eigenvalue grows exponentially like
$e^{\lambda t}$. For finite time $t$ the value of $\lambda$ depends
on initial conditions and it is called a finite time Lyapunov
exponent. The probability distribution of finite time Lyapunov
exponents $P(\lambda,t)$ \cite{ott,grassberger,tel}, especially its
behavior at small $\lambda$, is important for many applications,
where the measured quantity is sensitive to the existence of
trajectories staying close for anomalously long time. Examples range
from the problems of ocean acoustics~\cite{ocean} and branching of
2d electron flow~\cite{Top01} to Loschmidt echo~\cite{jalabert} and
mesoscopic superconductivity~\cite{Andreev}.

An evolution of small areas in phase space in the linearized
approximation is described by a combination of area preserving
stretching and squeezing (we assume two canonical variables $x$ and
$p$). In spite of a broad literature in the field the research which
goes beyond the linearized stability analysis of dynamics of close
trajectories is sparse. For dissipative systems it was done almost
twenty years ago by Politi {\it et al.}\cite{Henon1} and by Ott {\it
et al}.\cite{Henon2}. In those papers the authors considered the
consequences of tangency between stable and unstable manifolds for
chaotic attractors of nonhyperbolic two-dimensional maps. We are not
aware that similar analysis was applied to the Hamiltonian systems.
In this letter we try to fill this gap and consider the effect of
bending of phase space on $P(\lambda,t)$ for area preserving
systems.  Although our approach is similar to the approach of refs.
\cite{Henon1,Henon2}, their results were derived for the fractal
dimension spectrum rather than for the finite time Lyapunov
spectrum. Both characteristics are related to each other for
dissipative systems (see Grassberger {\it et al.} in
ref.\cite{grassberger}). However for Hamiltonian systems only the
Lyapunov spectrum is meaningful. Its behavior at small $\lambda$,
which is important for various practical
applications~\cite{ocean,Top01,jalabert,Andreev}, remains not
covered in the existing literature.
% and
%important for practical
%applications~\cite{ocean,Top01,jalabert,Andreev} features of its
%small $\lambda$ behavior remains not covered in the existing
%literature.
%more appropriate characteristic in Hamiltonian systems
%that naturally appears in various applications
%\cite{ocean,Top01,jalabert,Andreev}.

As we will see, a creation of narrow folds in the course of mapping
of the phase plane results in a great enhancement of the probability
to find small values of $\lambda$. As a result all negative moments
$\langle e^{-m\lambda t}\rangle$ for sufficiently large times %$t$
are saturated by the rare events with $\lambda\rightarrow 0$.

For large times the probability distribution of finite-time Lyapunov
exponents $P(\lambda,t)$ \cite{ott,grassberger,tel} has a generic
form ($\lambda\ge 0$)
 \bq\label{distribution}
P(\lambda,t)\sim \exp\left[tF(\lambda)\right],
 \ee
with a model specific function $F(\lambda)$ called
spectrum~\cite{grassberger}. The conventional selfaveraging Lyapunov
exponent $\lambda_0$ appears in the expansion of spectrum around the
maximum $F\approx -(\lambda-\lambda_0)^2/2\lambda_2$ (obviously
$\langle \lambda\rangle=\lambda_0$, ${\rm var}\lambda= \lambda_2/t$
for $\lambda_0 t\gg 1$). If the evolution of the system is described
by a Markovian process, i.e. if $P(\lambda, t)$ for large $t$ is
given by a convolution of $P(\lambda, t_i)$ for smaller time
intervals, then naturally $F(0)=-\infty$. On the other hand, in case
of mixed phase space one has $F(0)=0$, due to the possibility for
chaotic orbits to get "stuck" on the KAM
tori~\cite{grassberger,Zaslavsky}. In this letter we consider the
distribution of Lyapunov exponents in the
%principally
different regime of fully developed dynamical chaos (\emph{i.e}.
in the absence of stable periodic orbits). We argue that in this
case the spectrum still has a finite limit $F(0)={\rm const}<0$.

\begin{figure}
\includegraphics[width=7.cm, height=!]{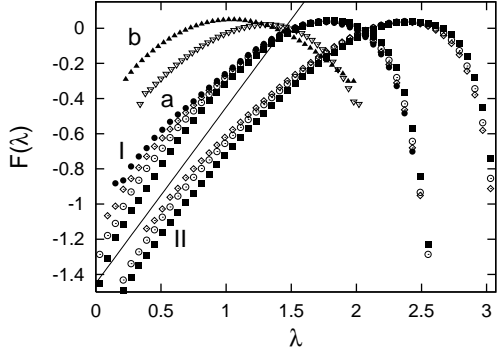}
%\twofigures{FFIG1.eps}{FFIG2.eps}
 \caption{ The finite-time
spectrum $F_t(\lambda)$ for the kicked rotator for different kicking
strengths $K=11$(I) and $K=20$(II), and times $t=10({\scriptstyle
\blacksquare}) , 15(\circ) ,20(\diamond) ,25(\bullet)$, and for the
time dependent random potential for two regimes $a$($\triangledown$)
and $b$($\blacktriangle$) described in the text. In case $b$ due to
the small values of Lyapunov exponent both $F_t(\lambda)$ and
$\lambda$ were multiplied by 5. The solid line shows a slope
$F'_t(\lambda) =1$. }
 \label{f.1}
\end{figure}
\begin{figure}
\includegraphics[width=7.cm, height=!]{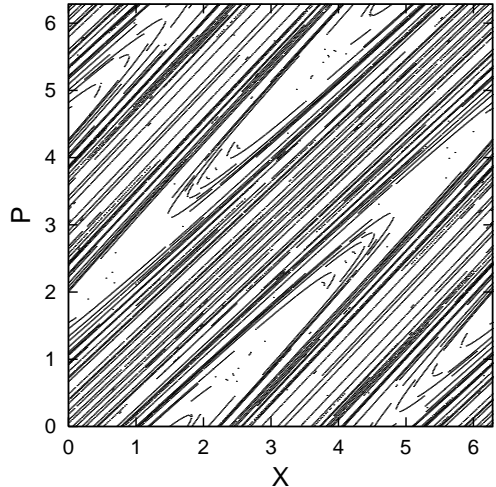}
 \caption{ Distribution of the endpoints of
trajectories $(x_t,p_t)$ having a small value of LE $\lambda <
\lambda_0/18, \lambda_0=\ln(K/2)$ for the kicked rotator with $K=20$
after 5 iterations. }
 \label{f.2}
\end{figure}

Dynamics of close trajectories is determined by the stability matrix
via
\begin{equation}
\left(\begin{array}{cc}
\delta p\\
\delta x
\end{array}\right)
={\cal M} \left(\begin{array}{cc}
\delta p_0\\
\delta x_0
\end{array}\right);
{\cal M}= \left(\begin{array}{cc}
\fr{\partial p}{\partial p_0}&\fr{\partial p}{\partial x_0}\\
\fr{\partial x}{\partial p_0}&\fr{\partial x}{\partial x_0}
\end{array}\right)
.\label{calMdef}
\end{equation}
The finite time Lyapunov exponent, specific for a given trajectory,
is defined as~\cite{endlambda}
 \bq\label{lambdadef}
\lambda=(2t)^{-1}\ln \left[\mbox{\rm tr} {\cal M}^T{\cal M}/2+\sqrt{
(\mbox{\rm tr} {\cal M}^T{\cal M}/2)^2-1}\right].
 \ee
Before presenting  the analytical theory, we show in Fig.~1 the
function $F_t(\lambda)=\ln(P(\lambda,t))/t$, found numerically for
two models. The first model, which we will primarily use, is the
kicked rotator~\cite{Chirikov79}:
 \begin{equation}\label{kickrot}
p_{n+1}=p_n+K\sin x_n,\quad x_{n+1}=x_n+p_{n+1}.
 \end{equation}
The spectrum $F(\lambda)$ in Eq.~(\ref{distribution}) is strictly
defined in the limit $t\rightarrow \infty$. Due to $\lambda\ge 0$,
its finite time approximations $F_t(\lambda)$ always have
$F_t(0)=-\infty$, but this singularity becomes weaker as time
increases. This tendency can be observed in Fig.~1, where different
finite time plots are presented (see especially the case for
$K=11$). Data on the figure includes $\sim 10^{12}$ trajectories.
Calculation of the spectrum for longer times would require the
unrealistic increase of the statistics. However the exponential
divergency of close trajectories is  already very noticeable, being
$\delta x(t)/ \delta x(0) \approx e^{45}$ for largest times in
Fig.~1.

The second model describes a particle moving under the action of a
time dependent random force, which we introduced via the ``random
kick'' Hamiltonian
 \bq
H=\fr{p^2}{2}+\kappa\sum_n\delta(t-n)\sum_{q=1}^N(a_{q}^{(n)}\sin qx
+b_{q}^{(n)}\cos qx)
 \ee
 with random coefficients $|a_{q}^{(n)}|,\ |b_{q}^{(n)}|<1$.
 Both plots in Fig.~1 are for $\kappa =1$: the plot~($a$) has
$N=3$ and 50 random kicks, while the plot~($b$) has $N=1$ and 300
kicks. The behavior of the spectrum for large $\lambda$ is
completely different for the two models. However, a shape of
$F(\lambda)$ at small $\lambda$ appears to be surprisingly similar.
The results of all simulations are consistent with a linear increase
at $\lambda\ll \lambda_0$, i.e. $F(0)={\rm const}, F'(0)=1$.

The distribution of finite time Lyapunov exponents, Eq.
(\ref{distribution}), is a rough statistical characteristic which
ignores correlations of values of $\lambda$ between different
trajectories. Such correlations are seen in Fig.~2, where we show
the endpoints of trajectories, which have a small $\lambda$. The
points on the phase plane are not distributed uniformly, as it would
be for a random process, but form a pattern of narrow curved lines.
Such a structure can be explained by the mechanism, schematically
presented in Fig.~3. The figure shows a time evolution of small area
of the phase plane around some classical trajectory $\{x(\tau),
p(\tau)\}$.  This evolution can be conveniently described by the
transformation of a small piece of square lattice covering the area
(Fig.~3a). Few stages of such an evolution are shown. (Estimates,
analogous to that of Eqs.~(\ref{stage1}-\ref{Rho}) below, were used
in Refs.~\cite{Henon1,Henon2} in order to describe the fractal
dimension spectrum $f(\alpha)$ of chaotic attractor.)

\begin{figure}
\includegraphics[width=7.cm, height=!]{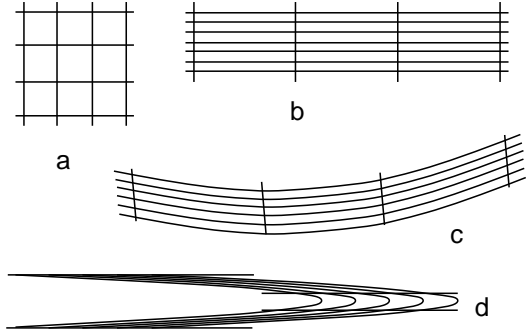}
%\onefigure[width=7.cm]{FFig03.eps}
 \caption{Schematic presentation
of the evolution of a small phase space area, described in the text.
} \label{f.3}
\end{figure}

The first stage, shown in Fig.~3b, consists of linear extension and
squeezing. After the appropriate orientation of coordinate frame
($x,p\rightarrow u,v$) it may be expressed as
 \bq\label{stage1}
u\rightarrow ue^{\lambda_0 t_1} \ , \ v\rightarrow ve^{-\lambda_0
t_1}.
 \ee
At this moment we ignore the fluctuations of the Lyapunov exponents
and use $\lambda_0$ for estimates. The linear transformation
(\ref{stage1}) always describes the evolution of infinitesimally
small areas. For small but finite regions of phase space a nonlinear
term should be added to Eq.~(\ref{stage1})
 \bq\label{stage2}
u \rightarrow u \ , \ v \rightarrow v+{u^2}/{2} ,
 \ee
 which is shown in Fig.~3c (the orientation of initial axes (a)
is different for Figs.~3b,c and d). Parabolas in Fig.~3c have a
curvature comparable to the size of the system, or the typical
momentum, both assumed to be of order unity. A consequent evolution
of the area consists of further stretching and squeezing. However,
the map, built from straight lines and parabolas shown in Fig.~3c,
is robust under this transformation. Squeezing of a parabola gives
another, more narrow, parabola. It is shown in Fig.~3d and may be
expressed as
 \bq\label{stage3}
u\rightarrow ue^{-\lambda_0 t_2} \ , \ v\rightarrow ve^{\lambda_0
t_2} .
 \ee
Since the orientation of unstable manifold changes by $\pi$ along
parabola, the squeezing always leads to diminishing of the
exponential divergency in some region near the vertex of the
parabola. Thus a stripe with small values of $\lambda$ is formed
along the centres of a set of narrow parabolas shown in Fig.~3d. The
curves in Fig.~2 show these stripes of small $\lambda$. Combining
together eqs.~(\ref{stage1}-\ref{stage3}) we express new coordinates
$u$ and $v$ through initial ones $u_0$ and $v_0$
 \bq\label{uv3}
u=u_0e^{\lambda_0 (t_1-t_2)}\ ,\ v=v_0e^{\lambda_0
(t_2-t_1)}+\fr{u_0^2}{2} e^{\lambda_0 (t_2+2t_1)}.
 \ee
Now the Lyapunov exponent is calculated with the help of the
stability matrix ${\cal M}$ via $\lambda \approx \ln {\cal
M}_{ij}^2/2t$~(\ref{lambdadef}):
 \bq\label{lambda}
\lambda= \fr{1}{2t}\ln \left[ 2{\rm ch}(2\lambda_0(t_1-t_2)) +
u^2e^{2\lambda_0 (t_2+t)} \right] ,
 \ee
where $t= t_1+t_2$. Thus the finite time Lyapunov exponent depends
only on one variable $u$. The function $\lambda(u)$~(\ref{lambda})
has a narrow minimum with the depth $ \lambda_{min}=\lambda_0
{|t_1-t_2|}/{t}$, and the width $\delta
u=\exp(\lambda_0(|t_1-t_2|-t- t_2))$. This explains the pattern seen
in Fig.~2. Away from the minimum we write
 \bq\label{lx}
\lambda=\lambda_0+(\ln |u|+\lambda_0t_2)/t \ \ \mbox{\rm or} \ \
|u|=e^{(\lambda-\lambda_0)t-\lambda_0 t_2}.
 \ee
 We see that at the
time $t$ small values of $\lambda$ are found in narrow stripes on
the $x,p$ plane, each corresponding to the set of a squeezed
parabolas (Fig.~3d). Among these lines those created at $t_1\approx
t/2$ have values as small as $\lambda\approx 1/t$, leading to
nonvanishing $P(\lambda,t)$ at the origin: $F(0)>-\infty$. In order
to estimate the probability distribution $P(\lambda,t)$ at
$\lambda\ll\lambda_0$, we first notice that a fold created at time
$t_1$ had a length $\sim 1$. It becomes as long as $e^{\lambda_0
t_2}$ at the time~$t$. Since the probability to find the Lyapunov
exponent in the interval $d\lambda$ is given by the area, where such
Lyapunov exponents are found, we may write
%This leads to
 \bq\label{Rho}
P(\lambda,t)\sim e^{\lambda_0 t_2}du/d\lambda \sim
\exp[(\lambda-\lambda_0)t].
 \ee
We keep only exponentially large contributions in this estimate.

\begin{figure}
%\onefigure{fig04.eps}
%\twofigures{fig04.eps}{fig05.eps}
\includegraphics[width=7.cm, height=!]{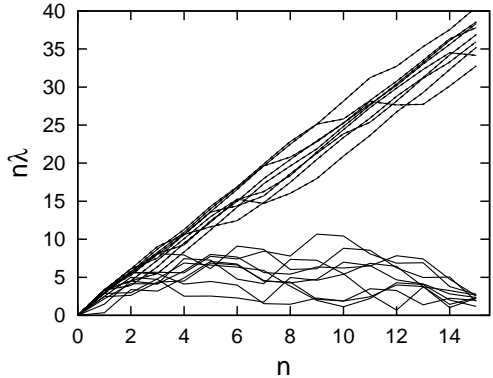}
% \twofigures{FFIG4.eps}{FFig05.eps}
 \caption{ $n\lambda$ versus $n$ for kicked
rotator with $K=20$. Examples of typical trajectories (slope
$\approx \ln(K/2)$, shown dashed), and rare trajectories having
$\lambda\le1/5$ after $15$ iterations (shown by solid lines). }
\label{f.4}
% \caption{The contour plot of the joint distribution
%$W(\lambda,\ln R)$ of the Lyapunov exponent and the logarithm of
%curvature, Eq.~(\ref{curvature}), for the kicked rotator with $K=20$
%after $15$ iterations. Contours were drawn by taking $10^{12}$
%initial conditions and allocating the results into the bins of
%two-dimensional histogram. The number of counts increases by $10$
%for each black contour, from outer to inner (the probability on the
%inner border of the contour is twice the probability on the outer
%side). The correlation between small $\lambda$ and small $R$
%confirms our phase-space-folding induced mechanism. }
% \label{f.5}
\end{figure}

\begin{figure}
\includegraphics[width=7.cm, height=!]{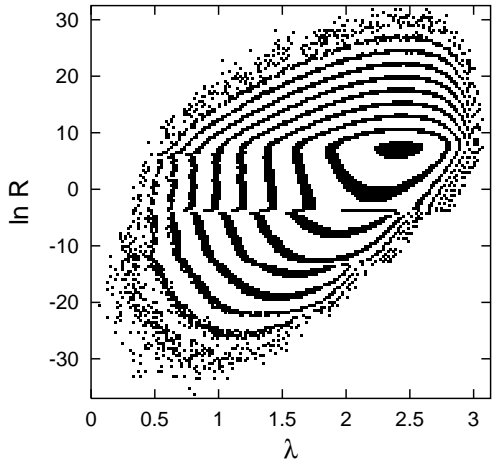}
 \caption{The contour plot of the joint distribution
$W(\lambda,\ln R)$ of the Lyapunov exponent and the logarithm of
curvature, Eq.~(\ref{curvature}), for the kicked rotator with $K=20$
after $15$ iterations. Contours were drawn by taking $10^{12}$
initial conditions and allocating the results into the bins of
two-dimensional histogram. The number of counts increases by $10$
for each black contour, from outer to inner (the probability on the
inner border of the contour is twice the probability on the outer
side). The correlation between small $\lambda$ and small $R$
confirms our phase-space-folding induced mechanism. }
 \label{f.5}
\end{figure}
In derivation of~(\ref{Rho}) we ignored fluctuations of Lyapunov
exponents during the squeezing stages (\ref{stage1}) and
(\ref{stage3}). Due to fluctuations $P(\lambda,t)$ is saturated by
only exponentially small fraction of strongest folds. Still we may
say that the simple folding mechanism described by the
eqs.~(\ref{stage1}-\ref{uv3}) is sufficient for having a finite
value of the spectrum at the origin $F(0)>-\infty$. It is also
capable to prevent the probability from decaying at small $\lambda$
faster than $P(\lambda,t)\propto e^{\lambda t}$. Thus we write
 \bq\label{asymspecw}
F(0)=-\tilde{\lambda}\gtrsim -\lambda_0 \ , \ F'(0) \le 1 .
 \ee

Taking into account fluctuations of the Lyapnov exponent during the
stages (\ref{stage1}) and (\ref{stage3}) may be done as
 \begin{eqnarray}
 \label{exact}
P(\lambda,t)&\sim& \int e^{\lambda t-\lambda_1 t_1-\lambda_2 t_2}
P(\lambda_1,t_1) P(\lambda_2,t_2)\nonumber\\
 &\times
&\delta(t_1+t_2-t)\delta(\lambda_1 t_1-\lambda_2 t_2-\lambda t)dt_1
dt_2 d\lambda_1 d\lambda_2 \ .
 \end{eqnarray}
Detailed investigation of this formula goes beyond the scope of this
paper. The behavior $P(\lambda,t)$ at small $\lambda$ may however be
analyzed easily. One readily shows, that the spectrum of the
form~(\ref{asymspecw}) with $F'\le 0$ is consistent with the
eq.~(\ref{exact}). Substitution of $P(\lambda,t)$ in the
form~(\ref{asymspecw}) into the r.h.s. of~(\ref{exact}) leads to the
same small $\lambda$ behavior in the l.h.s. This does not happen for
$F'(0)>1$. Also if $F'\le 0$, the integral over $\lambda_1$ and
$\lambda_2$ for small $\lambda$ is saturated by the region
$\lambda_1,\lambda_2 \ll \lambda_0$. Figure 4 shows
$n\lambda\sim\ln(\delta x_n/\delta x_0)$ for few typical
trajectories and for the rare trajectories with anomalously small
value of $\lambda$. As we see, typically two initially  very close
trajectories would escape from each other almost without
fluctuations $\delta x(t)\sim e^{\lambda_0 t}\delta x(0)$. For rare
trajectories, leading after large time $t$ to $\lambda(t)\approx 0$,
the extent of linear divergency is always much smaller, $\delta
x(t')\ll e^{\lambda_0 t'}\delta x(0)$ for any $t'<t$, in agreement
with~(\ref{exact}).

Although our result eq.~(\ref{asymspecw}) gives only a bound on how
fast the spectrum may increase with $\lambda$ , $F'(0)\le 1$,
numerical simulations of Fig.~1 are consistent with $F'(0)\approx
1$. Even in this "weak" form the result~(\ref{asymspecw}) has a
strong predictive power, leading to ($m=1,2,...$)
 \bq\label{moment}
t^{-1}\lim_{t\rightarrow\infty}\ln\langle e^{-m\lambda t}\rangle =
-\tilde{\lambda} .
 \ee
The dominant contribution to the averaged value here comes from the
rare trajectories with $e^{-\lambda t}\sim 1$. An exponentially
small value of $\langle e^{-m\lambda t}\rangle$ is due to a small
number of such trajectories. Eq.~(\ref{moment}) should be compared
with the often used Gaussian approximation $t^{-1}\ln\langle
e^{-m\lambda t}\rangle_{\rm G} \approx -m(\lambda_0-m\lambda_2/2) $.

Folding of the phase plane requires an introduction of a new
quantitative characteristic, in addition to Lyapunov exponent
$\lambda$ describing the linearized divergency of the trajectories.
To describe folding we introduce the radius of curvature $R$ of a
small area bended in the course of evolution. To this end, we first
analyze a \emph{nonlinear} divergency of the closed trajectories by
expanding the map~(\ref{kickrot}) to the second order in the
displacement between trajectories. If we write two components of the
displacement $\delta p, \delta x$ as a column ${\bf q}$, then
 \bq
{\bf q}_n={\cal M}_n{\bf q}_0+\fr{1}{2} \left(\begin{array}{cc}
1\\
0
\end{array}\right) {\bf q}_0^T
{\cal A}_n {\bf q}_0 +\fr{1}{2} \left(\begin{array}{cc}
0\\
1
\end{array}\right)
{\bf q}_0^T {\cal B}_n {\bf q}_0 ,
 \ee
where ${\cal M}_n, {\cal A}_n,{\cal B}_n$ are $2\times 2$
matrices. The stability matrix for kicked rotator is given by a
product
 \bq\label{stabil}
{\cal M}_n=\prod_{i<n}\left(\begin{array}{cc}
1 & K\cos x_i\\
1 & 1+K\cos x_i
\end{array}\right).
 \ee
Two other symmetric matrices may be found from the recursion
relations~\cite{areaconserv}.
 \begin{eqnarray}\label{stabilstabil}
&&{\cal A}_{n+1}={\cal A}_n+(1+K\cos x_n){\cal B}_n-K\sin x_n {\cal
C}_n , \nonumber\\
&&{\cal B}_{n+1}={\cal A}_{n+1}+{\cal B}_n ,
 \end{eqnarray}
where ${\cal A}_1={\cal B}_1=0$ and the elements of the matrix
${\cal C}_n$ are $({\cal C}_n)_{ij}=({\cal M}_n)_{i2}({\cal
M}_n)_{j2}$. Consider the radius of curvature $R$ of the image of a
line ${\bf q}_0^T\equiv(0, \delta x)$ as a measure of bending
(similar results may be obtained for the curvature of the image of
line ${\bf q}_0\equiv(\delta p,0)$). A simple calculation gives
 \bq
\fr{1}{R}=\fr{({\cal A}_n)_{22}({\cal M}_n)_{22}- ({\cal
B}_n)_{22}({\cal M}_n)_{12}}{(({\cal M}_n)_{12}^2+ ({\cal
M}_n)_{22}^2)^{3/2}}.\label{curvature}
 \ee
In order to analyze correlations between folding and the values of
the finite time Lyapunov exponent we introduce the joint
distribution $W(\lambda, \ln R)$. It is shown in
Fig.~5~\cite{footnote}. We see that for given $\lambda$ the
curvature may vary by orders of magnitude. However the general trend
is transparent: the smaller values of $\lambda$ correspond to the
smaller $R$~\cite{footnote1}.

The finite fraction of trajectories with $\lambda\approx 0$ in our
mechanism does not require the existence of any stable islands.
Still our main result (\ref{asymspecw}) includes also the case of
mixed phase space, corresponding to $\tilde{\lambda}=0$. In the
presence of very small stable islands one expects a competition of
two effects, when our mechanism will govern the small $\lambda$
spectrum at intermediate times, while the limit $t\rightarrow\infty$
will be due to dynamical traps~\cite{Zaslavsky} in the mixed phase
space.

In conclusion, for the case of developed chaos, the physical
quantities which may be calculated or measured naturally have a
statistical meaning. In this letter we have shown, how in such
systems folding of the $x,p$-plane inevitably leads to creation of
elongated areas, where initially diverging trajectories revert to
approaching to each other again. This allowed us to find a uniform
description of the tail of distribution of finite time Lyapunov
exponents, and to calculate all the negative moments of the
distribution~(\ref{moment}). Negative here means that we are able to
describe the physical quantities which are sensitive to the
contributions where, in-spite of chaos, trajectories stay longer
together and do not diverge.
%Since the physical quantities, such
%as coordinate and momentum, diverge exponentially with time
%increased the moments of this kind are most likely to be observed.

One immediate application of our findings will be a calculation of
the Loschmidt echo~\cite{jalabert}, which was shown~\cite{losch} to
decay on average like $\langle e^{-\lambda t}\rangle$. This behavior
should now be reconsidered in view of our eq.~(\ref{moment}).
Another problem of current interest, where the result is sensitive
to the existence of long trajectories with small $\lambda$, is a
calculation of the gap in the spectrum of the Andreev
billiard~\cite{Andreev}.  Application of our result
Eq.~(\ref{asymspecw}) to this problem leads to the strong
modification of the low energy Andreev spectrum and substantial
lowering of the superconducting gap~\cite{Silv05}.
%Calculation of
%the low energy Andreev spectrum due to large fluctuations of
%finite time Lyapunov exponents is currently in
%progress~\cite{Silv05}.

%\acknowledgments
Discussions with {C.~W.~J.~Beenakker}, A.~N.~Morozov,
{J.~Tworzyd\l o}, and J.~Whitaker are greatly appreciated. This
work was supported by the Dutch Science Foundation NWO/FOM and by
the SFB TR 12.

%\begin{thebibliography}{00}
% notes:
% \bibitem{label2}
% If there is a note, it should come last:
% \bibitem{label3} \note
% \end{subbibitems}
%\bibitem{}
%\end{thebibliography}


\begin{thebibliography}{99}
\bibitem{endlambda}
The matrix ${\cal {M}}$ may be decomposed as ${\cal {M}}={\cal
{U}}{\cal {T}}$, where a real symmetric matrix ${\cal T}$ ($\det
{\cal T}\equiv 1$) describe squeezing and stretching with respect to
some normal axes, and a normal matrix ${\cal U}$ describes the
rotation of the plane. We define $\lambda$~(\ref{lambdadef}) through
the eigenvalue of ${\cal{M}}^T{\cal{M}}={\cal{T}}^T{\cal{T}}$.
\bibitem{footnote}
The distribution $W(\lambda,\ln R)$ has a main peak at $R\sim K^2$
and $\lambda\approx\ln(K/2)$. Second narrow peak at $R\sim 1/K$
corresponds to a minimal curvature which can be found from
(\ref{curvature}) after a single iteration of the map
(\ref{kickrot}).
\bibitem{footnote1} This distribution differs drastically from what
we find for the mixed phase space. Mixed phase space takes place in
the same model (\ref{kickrot}), but with smaller values of $K$
($K\lesssim 7$). In this case we found an extra pronounced maximum
of $W(\lambda,\ln R)$ at small $\lambda$ and $R\approx 1$. Moreover,
even the exponentially small stability islands, which are hard to
observe via the Poincare section, modify significantly the bivariate
distribution $W(\lambda,\ln R)$.
\end{thebibliography}
\end{document}